\documentclass[twocolumn]{revtex4}

\usepackage{graphicx}
\usepackage{dcolumn}
\usepackage{amsmath}

\begin{document}

\title{Renormalization group theory of generalized multi-vertex sine-Gordon
model
}

\author{Takashi Yanagisawa
}

\affiliation{
Electronics and Photonics Research Institute,
National Institute of Advanced Industrial Science and Technology,
1-1-1 Umezono, Tsukuba, Ibaraki 305-8568, Japan
}

\begin{abstract}
We investigate the renormalization group theory of generalized multi-vertex
sine-Gordon model by employing the dimensional regularization method and
also the Wilson renormalization group method.
The vertex interaction is given by $\cos(k_j\cdot \phi)$ where 
$k_j$ ($j=1,2,\cdots,M$) are momentum  vectors and $\phi$ is an $N$-component
scalar field.
The beta functions are calculated for the sine-Gordon model with multi
cosine interactions.
The second-order correction in the renormalization procedure is given by the
two-point scattering amplitude for tachyon scattering. 
We show that new vertex interaction with momentum vector $k_{\ell}$ is generated 
from two vertex interactions with vectors $k_i$ and $k_j$ 
when $k_i$ and $k_j$ meet the condition  
$k_{\ell}=k_i\pm k_j$ called the triangle condition.
Further condition $k_i\cdot k_j=\pm 1/2$ is required
within the dimensional regularization method.
The renormalization group equations form a set of closed equations
when $\{k_j\}$ form an equilateral triangle for $N=2$ or a regular tetrahedron
for $N=3$.
The Wilsonian renormalization group method gives qualitatively the
same result for beta functions.
\end{abstract}


\maketitle

\section{Introduction}

The sine-Gordon model is an interesting universal model that appears
as an effective model in various fields of 
physics\cite{col75,das75,jos77,sam78,zam79,ami80,wie78,tse90,bal00,raj82,man04,
col85,mar17,wei15,her07,nan04,nag09}. 
The two-dimensional (2D) sine-Gordon model can be mapped to the
Coulomb gas model that has logarithmic Coulomb interaction\cite{jos76,zin89}. 
The 2D sine-Gordon model has been investigated by several methods, 
especially by using the renormalization group method.
The physics of the sine-Gordon model is closely related to that of 
the Kosterlitz-Thouless
transition of the 2D XY model\cite{kos73,kos74}.

The sine-Gordon model is the model of scalar field under the periodic
potential.
This model can be generalized in several ways.
The massive chiral model is regarded as a generalization of the 
sine-Gordon model where the potential term ${\rm Tr}(g+g^{-1})$ is 
considered for $g$ in a gauge group (Lie group) $G$ ($g\in G$)\cite{gol78,bre79,yan16}. 
The chiral model was generalized to include the Wess-Zumino term
as the Wess-Zumino-Witten model\cite{wes71,wit83,wit84,nov82}.
The other way of generalization is to include the potential terms of high 
frequency modes\cite{nan09,del98}.
A generalized potential term is given as
\begin{equation}
V = \frac{1}{t}\sum_{n=1}^L\alpha_n\cos(n\phi),
\end{equation}
where $\phi$ is a one-component scalar field and $L$ is an integer. 
In the Wilson renormalization group method, the cosine potential
$\cos((n-m)\phi)$ is generated from $\cos(n\phi)$ and $\cos(m\phi)$ in the 
second order perturbation.  Thus there will be the correction to the
beta function of $\alpha_n$ in the form $\alpha_{\ell}\alpha_m$ with $n=|\ell-m|$. 
For the hyperbolic sine-Gordon model, $\alpha_{n}$ has a correction
from $\alpha_{\ell}$ and $\alpha_m$ satisfying $n=\ell+m$\cite{yan19}.

This kind of model can be generalized to a multi-component scalar
field.
In this paper we investigate a generalized multi-component sine-Gordon
model with multiple cosine potentials.
The cosine vertex interaction is given by $\cos(\sum_{\ell}k_{j\ell}\phi_{\ell})$
where $\phi=(\phi_1,\cdots,\phi_N)$ is a scalar field and 
$k_j=(k_{j1},\cdots,k_{jN})$ ($j=1,\cdots,M$) are momentum vectors of real numbers.
$k_j$ represents the direction of oscillation of field $\phi$.
The model for $M=3$ was considered in \cite{oak19}.
The condition to generate a new vertex interaction shown above is
generalized to $k_n=k_{\ell}\pm k_m$.
This is called the triangle condition in this paper.

It was pointed out that there is a close relation between the sine-Gordon
model and string propagation in a tachyon background\cite{das86}.
In fact, two-vertex correction in the renormalization procedure is
given by the two-point scattering amplitude for tachyon scattering
in the second order perturbation theory.
The multi-vertex correction will be given by the multi-point tachyon
scattering amplitude.

This paper is organized as follows.
In section III we present the generalized sine-Gordon model.
We show the renormalization procedure based on the dimensional
regularization method in section IV.
We applied the Wilsonian renormalization group method to our model
in section V.  We consider the generalized multi-vertex sine-Gordon
model and calculate the beta functions in section VI.
We give summary in the last section.

\section{Multi-vertex sine-Gordon model}

We consider an $N$-component real scalar field 
$\phi= (\phi_1,\cdots,\phi_N)$.
The model is a $d$-dimensional generalized multi-vertex 
sine-Gordon model given by
\begin{equation}
\mathcal{L}= \frac{1}{2t_0}\left(\partial_{\mu}\phi\right)^2
+ \frac{1}{t_0}\sum_{j=1}^M\alpha_{0j}\cos\left(k_j\cdot\phi\right),
\end{equation}
where $t_0 (>0)$ and $\alpha_{0j}$ $(j=1,\cdots,M)$ are bare 
coupling constants and
$k_j$ $(j=1,\cdots,M)$ are $N$-component constant vectors.
We use the notation $(\partial_{\mu}\phi)^2=\sum_j(\partial_{\mu}\phi_j)^2$
and $k_j\cdot\phi=\sum_{\ell}k_{j\ell}\phi_{\ell}$ for
$k_j=(k_{j1},\cdots,k_{jN})$.
We use the Euclidean notation in this paper.
The second term is the potential energy with multi cosine
interactions.
The dimensions of $t_0$ and $\alpha_{0j}$ are given as
$[t_0]=\mu^{2-d}$ and $[\alpha_{0j}]=\mu^2$ for the energy scale
parameter $\mu$.
The analysis is performed near two dimensions $d=2$.
We introduce the renormalized coupling constants $t$ and
$\alpha_j$ where the renormalization constants are defined as
\begin{equation}
t_0= t\mu^{2-d}Z_t,~~ \alpha_{0j}= \alpha_j\mu^2Z_{\alpha j},
\end{equation}
where we set that $t$ and $\alpha_j$ are dimensionless constants.
The renormalized field $\phi_R$ is introduced with the
renormalization constant $Z_{\phi}$ as follows
\begin{equation}
\phi = \sqrt{Z_{\phi}}\phi_R.
\end{equation}
In the following $\phi$ denotes the renormalized field $\phi_R$
for simplicity.
Then the Lagrangian density is given as
\begin{equation}
\mathcal{L}= \frac{\mu^{d-2}Z_{\phi}}{2tZ_t}
\left(\partial_{\mu}\phi\right)^2
+\frac{\mu^d}{tZ_t}\sum_jZ_{\alpha_j}\alpha_j \cos(k_j\cdot\phi).
\end{equation}
We examine the renormalization group procedure for this model in section
III and section IV.
We also investigate the component dependence of renormalization in
section V by generalizing the model as follows.
\begin{equation}
\mathcal{L}= \sum_{\ell}\frac{\mu^{d-2}Z_{\phi}}{2t_{\ell} Z_{t_{\ell}}}
(\partial_{\mu}\phi_{\ell})^2
+\sum_j\frac{\mu^d\alpha_j Z_{\alpha_j}}{t_jZ_{t_j}}
\cos\left(\sqrt{Z_{\phi}}k_j\cdot\phi\right).
\end{equation} 

We need some conditions so that we have one fixed point for $t$.
For this purpose we normalize $k$ vectors as
\begin{equation}
k_j^2 = \sum_{\ell=1}^{N}k_{j\ell}^2= 1 ~~(j=1,\cdots,M).
\end{equation}
From the two vertices with momentum vectors $k_i$ and $k_j$, new 
vertex is generated
with momentum $k_m$ when the triangle condition is satisfied:
\begin{equation}
k_m= k_i\pm k_j.
\end{equation}  
We assume that a set $\{\alpha_j\}$
includes all vertices that will be generated from multi-vertex interactions
each other.
For a triangle or a regular polyhedron which is composed of
equilateral triangles, $M$ becomes finite since $\{k_j\}$ form
a finite set.
For example, we consider an equilateral triangle
or a regular tetrahedron. 
For an equilateral triangle ($N=2$, $M=3$) or a regular
tetrahedron ($N=3$, $M=6$), we have
\begin{equation}
\sum_{j=1}^Mk_{j\ell}^2 = C(M) ~~~ {\rm for}~\ell=1,\cdots,N,
\end{equation}
where $C(M)$ is a constant depending upon $M$.
These conditions will be explained in the following sections.

\section{Renormalization by dimensional regularization}

We evaluate the beta functions for the multi-vertex sine-Gordon
model by using the dimensional regularization method\cite{tho72,gro76,yan17}.

\subsection{Tadpole renormalization of $\alpha_j$}

The lowest order contributions to the renormalization of $\alpha_j$
are given by tadpole diagrams.
Using the expansion $\cos\phi=1-\frac{1}{2}\phi^2+\frac{1}{4!}\phi^4-\cdots$,
the cosine potential is renormalized as
\begin{equation}
\cos(\sqrt{Z_{\phi}}k_j\cdot\phi)~\rightarrow ~\left( 1-\frac{1}{2}Z_{\phi}
\langle(k_j\cdot\phi)^2\rangle+\cdots\right)\cos(\sqrt{Z_{\phi}}k_j\cdot\phi).
\end{equation}
$\langle\phi^2\rangle$ is regularized as
\begin{equation}
\langle\phi_{\ell}^2\rangle = \frac{t\mu^{2-d}Z_t}{Z_{\phi}}\int
\frac{d^dk}{(2\pi)^d}\frac{1}{k^2+m_0^2}
= -\frac{t\mu^{2-d}Z_t}{Z_{\phi}}\frac{1}{\epsilon}
\frac{\Omega_d}{(2\pi)^d},
\end{equation} 
for $d=2+\epsilon$ where a mass $m_0$ is introduced to avoid
the infrared divergence.
We set $Z_t=1$ in the lowest order of $t$.
We adopt that $\langle\phi_i\phi_{\ell}\rangle=\delta_{i\ell}\langle\phi_i^2\rangle$
and $\langle\phi_{\ell}^2\rangle$ is independent of $\ell$.
Then the renormalization of the potential term is given as
\begin{eqnarray}
&& \alpha_j Z_{\alpha j}\left( 1-\frac{1}{2}Z_{\phi}k_j^2
\langle\phi_1^2\rangle+\cdots \right)
\cos\left(\sqrt{Z_{\phi}}k_{j}\cdot\phi\right)\nonumber\\
&&~~~ = \alpha_j Z_{\alpha j}\left( 1+\frac{1}{2}k_j^2\frac{1}{\epsilon}
t\mu^{2-d}\frac{\Omega_d}{(2\pi)^d}+\cdots \right)\nonumber\\
&& ~~~~~ \cdot \cos\left(\sqrt{Z_{\phi}}k_{j}\cdot\phi\right).
\end{eqnarray} 
The renormalization constant $Z_{\alpha j}$ is determined as
\begin{equation}
Z_{\alpha j}= 1-\frac{t}{4\pi\epsilon}k_j^2.
\end{equation}
Since the bare coupling constant $\alpha_{0j}=\alpha_j\mu^2 Z_{\alpha j}$ is
independent of $\mu$, we have 
\begin{equation}
\beta(\alpha_j):= \mu\frac{\partial \alpha_j}{\partial\mu}= -2\alpha_j
+\frac{1}{4\pi}k_j^2 t\alpha_j.
\end{equation}
The beta function of 
$\alpha_j$ has a zero at
\begin{equation}
t=t_{cj}= 8\pi/k_j^2=8\pi,
\end{equation}
since $k_j^2=1$.

\subsection{Vertex-vertex interaction}
 We investigate the corrections to $t$ and $\alpha_j$ from
vertex-vertex interactions.
The second-order contribution $I^{(2)}$ to the action is given by
\begin{eqnarray}
I^{(2)}&=& -\frac{1}{2}\left(\frac{\mu^d}{tZ_t}\right)^2 \int d^dx d^dx'
\nonumber\\
&\times& \sum_{ij}\alpha_i \alpha_j Z_{\alpha_i}Z_{\alpha_j} 
\cos\left(\sqrt{Z_{\phi}}k_i\cdot \phi(x)\right) \nonumber\\
&& \times \cos\left( \sqrt{Z_{\phi}}k_j\cdot \phi(x')\right) \nonumber\\
&=& -\frac{1}{4}\left(\frac{\mu^d}{tZ_t}\right)^2 \int d^dx d^dx'
\nonumber\\
&\times& \sum_{ij}\alpha_i \alpha_j Z_{\alpha_i}Z_{\alpha_j} 
\bigg[ \cos\left(\sqrt{Z_{\phi}}(k_i\cdot\phi(x)-k_j\cdot\phi(x'))\right) 
\nonumber\\
&& +\cos\left(\sqrt{Z_{\phi}}(k_i\cdot\phi(x)+k_j\cdot\phi(x'))\right)
\bigg]. 
\end{eqnarray}

We first examine the first term denoted as $I_1^{(2)}$:
\begin{eqnarray}
I_1^{(2)}&=& -\frac{1}{4}\left(\frac{\mu^d}{tZ_t}\right)^2 \int d^dx d^dx'
\nonumber\\
&\times& \sum_{ij}\alpha_i \alpha_j Z_{\alpha_i}Z_{\alpha_j} 
\cos\left(\sqrt{Z_{\phi}}(k_i\cdot\phi(x)-k_j\cdot\phi(x'))\right).
\nonumber\\
\end{eqnarray} 
We evaluate the renormalization of cosine term by calculating
$\langle (k_i\cdot\phi-k_j\cdot\phi)^2\rangle$.
We adopt that 
$\langle \phi_{\ell}(x)\phi_m(x')\rangle=
\delta_{\ell m}\langle\phi_{\ell}(x)\phi_{\ell}(x')\rangle$, and
$\langle\phi_{\ell}(x)\phi_{\ell}(x')\rangle$ is independent of $\ell$:
$\langle\phi_{\ell}(x)\phi_{\ell}(x')\rangle=\langle\phi_1(x)\phi_1(x')\rangle$.
Then
\begin{eqnarray}
&& \langle(k_i\cdot\phi-k_j\cdot\phi)^2\rangle \nonumber\\
&=& \sum_{\ell}\bigg[ k_{i\ell}^2\langle\phi_{\ell}(x)^2\rangle
+k_{j\ell}^2\langle\phi_{\ell}(x')^2\rangle
-2k_{i\ell}k_{j\ell}\langle\phi_{\ell}(x)\phi_{\ell}(x')\rangle \bigg].
\nonumber\\
\end{eqnarray}
$I_1^{(2)}$ is renormalized as
\begin{eqnarray}
I_1^{(2)} &=& -\frac{1}{4}\left(\frac{\mu^d}{tZ_t}\right)^2
\int d^dx d^dx' \bigg[ \sum_i\alpha_i^2 Z_{\alpha_i}^2 \nonumber\\
&& \times \exp\left( -Z_{\phi}k_i^2\langle\phi_1(x)^2\rangle
+Z_{\phi}k_i^2\langle\phi_1(x)\phi_1(x')\rangle \right) \nonumber\\
&& \times \cos\left( \sqrt{Z_{\phi}}(k_i\cdot (\phi(x)-\phi(x'))) \right)
\nonumber\\
&+& \sum_{i\neq j}\alpha_i\alpha_j Z_{\alpha_i}Z_{\alpha_j}
\exp\Big[ -\frac{Z_{\phi}}{2}(k_i^2\langle\phi_1(x)^2\rangle \nonumber\\ 
&& +k_j^2\langle\phi_1(x')^2\rangle )  
 +Z_{\phi}k_i\cdot k_j\langle\phi_1(x)\phi_1(x')\rangle \Big]\nonumber\\
&&\times \cos\left(\sqrt{Z_{\phi}}(k_i\cdot \phi(x)-k_j\cdot\phi(x'))\right) \bigg].
\end{eqnarray} 
The two-point function is written as
\begin{eqnarray}
\langle\phi_{\ell}(x)\phi_{\ell}(y)\rangle &=&
\frac{t\mu^{2-d}Z_t}{Z_{\phi}}\int\frac{d^dp}{(2\pi)^d}
\frac{e^{ip\cdot (x-y)}}{p^2+m_0^2} \nonumber\\
&=& \frac{t\mu^{2-d}Z_t}{Z_{\phi}}\frac{\Omega_d}{(2\pi)^d}
K_0(m_0|x-y|),
\end{eqnarray}
where we introduced $m_0$ to avoid the infrared divergence and
$K_0$ is the zero-th modified Bessel function.

\subsection{Renormalization of $t$}

The first term of $I_1^{(2)}$ gives a contribution to the renormalization of the
coupling constant $t$.
Since $K_0(m_0r)$ increases as $r\rightarrow 0$, we can expand in terms of $r$.
By using 
$\cos(\sqrt{Z_{\phi}}k_i\cdot (\phi(x)-\phi(x+{\bf r})))\simeq 
1-(1/2)Z_{\phi}(r_{\mu}\partial_{\mu}(k_i\cdot \phi(x)))^2$ where
$\partial_{\mu}=\partial/\partial x_{\mu}$,  
the first term $I_{1a}^{(2)}$ of $I_1^{(2)}$ is written as
\begin{eqnarray}
I_{1a}^{(2)} &\simeq& \frac{1}{4}\left(\frac{\mu^d}{tZ_t}\right)^2
\int d^dx d^dr \sum_i\alpha_i^2 Z_{\alpha_i}^2
\frac{1}{4}Z_{\phi}(\partial_{\mu}\tilde{\phi_i})^2 \nonumber\\
&\times& r^2\exp\left( -Z_{\phi}k_i^2\langle\phi_1^2\rangle
+tZ_tk_i^2\frac{\Omega_d}{(2\pi)^d}K_0(m_0r) \right), \nonumber\\
\end{eqnarray}
where $\tilde{\phi_i}=\sum_{\ell}k_{i\ell}\phi_{\ell}$. 
If $(k_{j\ell})\in SO(N)$ (with $M=N$), we have
$\sum_i(\partial_{\mu}\phi_i)^2=\sum_i(\partial_{\mu}\tilde{\phi_i})^2$.
In general, we have
\begin{eqnarray}
\sum_i(\partial_{\mu}\tilde{\phi_i})^2 &=& \sum_{i\ell}k_{i\ell}^2
(\partial_{\mu}\phi_{\ell})^2
+\sum_{i,\ell\neq m}k_{i\ell}k_{im}\partial_{\mu}\phi_{\ell}\partial_{\mu}\phi_m.
\nonumber\\
\end{eqnarray}
As mentioned in section II, we consider the case where $\{k_j\}$ form an equilateral
triangle ($M=3$) or a regular tetrahedron ($M=6$), and we obtain
$\sum_ik_{i\ell}^2= {\rm constant} \equiv C$ depending on $M$ such as
$C=3/2$ for $M=3$ and $N=2$, and $C=2$ for $M=6$ and $N=3$.
In this case
\begin{equation}
\sum_i(\partial_{\mu}\tilde{\phi_i})^2 = 
C\sum_i(\partial_{\mu}\phi_i)^2
+\sum_{i,\ell\neq m}k_{i\ell}k_{im}\partial_{\mu}\phi_{\ell}\partial_{\mu}\phi_m.
\end{equation}
Inorder to recover the kinetic term in the original action, we use
the approximation
\begin{eqnarray}
\sum_i\alpha_i^2Z_{\alpha_i}^2(\partial_{\mu}\tilde{\phi_i})^2
&\rightarrow& \frac{1}{M}\sum_i\alpha_i^2\cdot C(\partial_{\mu}\phi)^2
\nonumber\\
&\equiv& \langle\alpha_i^2\rangle C(\partial_{\mu}\phi)^2.
\end{eqnarray}
Otherwise the renormalization of the kinetic term becomes complicated
since we must introduce $\{t_i\}$ that depend on components of $\phi$.
This may not be essential for the renormalization group flow.
We discuss this point later.

$\langle\phi_1^2\rangle$ is evaluated as
\begin{equation}
\langle\phi_1^2\rangle= \frac{t\mu^{2-d}Z_t}{Z_{\phi}}
\frac{\Omega_d}{(2\pi)^d}K_0(m_0a),
\end{equation}
where $a$ is a small cutoff.
The $r$-integration is calculated as
\begin{eqnarray}
J_j &:=& \int d^dr r^2\exp\left( tk_j^2\frac{\Omega_d}{(2\pi)^d}
K_0(m_0\sqrt{r^2+a^2})\right) \nonumber\\
&\simeq& \Omega_d \int dr r^{d+1}\left(
\frac{1}{cm_0^2(r^2+a^2)}\right)^{\frac{t}{4\pi}k_j^2},
\end{eqnarray}
where $c=(e^{\gamma}/2)^2$. We put
\begin{equation}
d= 2+\epsilon,
\end{equation}
and 
\begin{equation}
\frac{t}{8\pi} = 1+v,
\end{equation}
since we normalize $k_j^2=1$.
Then we have
\begin{eqnarray}
J_j&=& \Omega_s(cm_0^2)^{-2}\int_0^{\infty}dr r^{d+1}
\frac{1}{(r^2+a^2)^{2+2v}} \nonumber\\
&=& -\Omega_d(cm_0^2)^{-2}\frac{1}{\epsilon}+O(v).
\end{eqnarray}
This indicates
\begin{eqnarray}
I_{1a}^{(2)}&=& -\frac{C}{8}\left(\frac{\mu^d}{tZ_t}\right)^2
\langle\alpha_i^2\rangle\exp\left(-Z_{\phi}
\langle\phi_1^2\rangle\right) \nonumber\\
&& \times \Omega_d (cm_0^2)^{-2}\frac{1}{\epsilon}\int d^dx
\frac{1}{2}Z_{\phi}k_i^2(\partial_{\mu}\phi)^2 + O(v_i).
\nonumber\\
&=& -\frac{C}{8}\left(\frac{\mu^d}{tZ_t}\right)^2
\langle\alpha_i^2\rangle(cm_0^2a^2)^{tZ_t/4\pi}
\Omega_d(cm_0^2)^{-2}\nonumber\\
&& \times \frac{1}{\epsilon}\int d^dx\frac{1}{2}Z_{\phi}
(\partial_{\mu}\phi)^2+O(v)\nonumber\\
&=& -\frac{C}{8}\langle\alpha_i^2\rangle\frac{\Omega_d}{8\pi}
\mu^{d+2}a^4\frac{1}{\epsilon} \int d^dx
\frac{\mu^{d-2}Z_{\phi}}{2tZ_t}(\partial_{\mu}\phi)^2 \nonumber\\
&& +O(v).
\end{eqnarray}
Then we choose
\begin{equation}
Z_t = 1-\frac{C}{32}\langle\alpha_i^2\rangle\mu^{d+2}a^4\frac{1}{\epsilon},
\end{equation}
where we set $Z_{\alpha_i}=1$ to the lowest order of $\alpha_i$.

Since the bare coupling constant $t_0=t\mu^{2-d}Z_t$ is independent
of the energy scale $\mu$, we have $\mu\partial t_0/\partial\mu=0$.
This results in
\begin{eqnarray}
\beta(t) &:=& \mu\frac{\partial t}{\partial \mu}
= (d-2)t-t\mu\frac{\partial \ln Z_t}{\partial\mu} \nonumber\\
&=& (d-2)t+\frac{C}{32}t\langle\alpha_i^2\rangle,
\end{eqnarray}
where we used $\mu\partial\alpha_i/\partial\mu=-2(\alpha_i-t/8\pi)$,
neglecting terms of the order of $t^2\alpha_i^2$,
and we put $a=\mu^{-1}$.

\subsection{Vertex-vertex correction to $\alpha_j$}

We consider the second term in $I_{1b}^{(2)}$ that contains
multi-vertex interaction:
\begin{eqnarray}
I_{1b}^{(2)} &:=& -\frac{1}{4}\left(\frac{\mu^d}{tZ_t}\right)^2
\int d^dx d^dx' 
\sum_{i\neq j}\alpha_i\alpha_j Z_{\alpha_i}Z_{\alpha_j} \nonumber\\
&& \times \exp\Big[ -\frac{Z_{\phi}}{2}\left(k_i^2\langle\phi_1(x)^2\rangle 
 +k_j^2\langle\phi_1(x')^2\rangle \right) \nonumber\\ 
&& +Z_{\phi}k_i\cdot k_j\langle\phi_1(x)\phi_1(x')\rangle \Big]\nonumber\\
&&\times \cos\left(\sqrt{Z_{\phi}}(k_i\cdot \phi(x)-k_j\cdot\phi(x'))\right).
\end{eqnarray} 
Let us examine the integral given by
\begin{eqnarray}
J_{ij} &:=& \int d^dr\exp\left( Z_{\phi}k_i\cdot k_j\langle\phi(x)_1\phi_1(x+r)
\rangle \right) \nonumber\\
&=& \int d^dr \exp\left(k_i\cdot k_j t\mu^{2-d}Z_t
\frac{\Omega_d}{(2\pi)^d}K_0(m_0 r) \right) \nonumber\\
&\simeq& \Omega_d\int_0^{\infty}dr r^{d-1}
\left(\frac{1}{cm_0^2(r^2+a^2)}\right)^{tk_i\cdot k_j/4\pi} \nonumber\\
&=& \Omega_d\left(\frac{1}{cm_0^2a^2}\right)^{tk_i\cdot k_j}\frac{a^{d/2}}{2}
\frac{1}{\Gamma(tk_i\cdot k_j/4\pi)}\Gamma\left(\frac{d}{2}\right)
\nonumber\\
&& \times \Gamma\left( \frac{t}{4\pi}k_i\cdot k_j-\frac{d}{2}\right),
\end{eqnarray}
where the cutoff $a$ is introduced.
We put $t=8\pi(1+v)$, then we have a divergence near two dimensions when
\begin{equation}
k_i\cdot k_j= 1/2.
\end{equation}
This means that two vectors $k_i$ and $k_j$ forms an equilateral
triangle.   When $k_i$ and $k_j$ satisfy this condition, we have
\begin{equation}
J_{ij}= -\Omega_d (cm_0^2)^{-1}\frac{1}{\epsilon}+O(v).
\end{equation}
Then we obtain
\begin{eqnarray}
I_{1b}^{(2)}&\simeq& \frac{1}{4}\left(\frac{\mu^d}{tZ_t}\right)^2
\sum_{i\neq j}\alpha_iZ_{\alpha_i}\alpha_jZ_{\alpha j}
\left(cm_0^2a^2\right)^{t/4\pi} \nonumber\\
&&\times \frac{1}{\epsilon}\Omega_d(cm_0^2)^{-1}\int d^dx
\cos\left( \sqrt{Z_{\phi}}(k_i-k_j)\cdot\phi(x) \right) \nonumber\\
&\simeq& \frac{1}{\epsilon}\frac{cm_0^2}{16}\sum_{i\neq j}\alpha_i
\alpha_j\frac{1}{tZ_t}\mu^{2d}a^4 \int d^dx \nonumber\\
&&\times \cos\left( \sqrt{Z_{\phi}}(k_i-k_j)\cdot\phi(x) \right).
\end{eqnarray}
Let $k_{\ell}$ be a vector so that $k_i$, $k_j$ and $k_{\ell}$ form
an equilateral triangle where 
\begin{equation}
k_i-k_j=k_{\ell}.  
\end{equation}
Then the
potential term with coefficient $\alpha_{\ell}$ has the correction as
\begin{equation}
\frac{\mu^d}{tZ_t}\alpha_{\ell}Z_{\alpha_{\ell}}
\left(1+\frac{t}{4\pi\epsilon}+\frac{1}{16\epsilon}
\frac{\alpha_i\alpha_j}{\alpha_{\ell}}\mu^dcm_0^2a^4 \right) 
\cos\left(\sqrt{Z_{\phi}}k_{\ell}\cdot\phi\right).
\end{equation}
We choose the renormalization constant $Z_{\alpha_{\ell}}$ as
\begin{equation}
Z_{\alpha_{\ell}}= 1-\frac{t}{4\pi\epsilon}
-\frac{1}{16\epsilon}\frac{\alpha_i\alpha_j}{\alpha_{\ell}}\mu^dcm_0^2a^4.
\end{equation}
This leads to the beta function $\beta(\alpha_{\ell})$ with correction as
\begin{equation}
\beta(\alpha_{\ell})= -2\alpha_{\ell}\left(1-\frac{t}{8\pi}\right)
+\frac{1}{8}cm_0^2a^2\alpha_i\alpha_j.
\end{equation}
Since the coefficient of the correction term is dependent on cutoff parameters,  
we choose $cm_0^2a^2=1$ to have
\begin{equation}
\beta(\alpha_{\ell})= -2\alpha_{\ell}\left(1-\frac{t}{8\pi}\right)
+\frac{1}{16}\alpha_i\alpha_j.
\label{b-alpha}
\end{equation}

There is also a contribution from the second term in $I^{(2)}$ where
$k_j$ is replaced by $-k_j$.  In this case vertices with $k_i\cdot k_j=-1/2$
generate a new vertex with $k_{\ell}$ satisfying
\begin{equation}
k_i+k_j= k_{\ell}.
\end{equation}
In the dimensional regularization method, two vertices satisfying
$k_i\cdot k_j=\pm 1/2$ generate a new vertex with $k_i\mp k_j$.
As a result, the beta function for $\alpha_{\ell}$ reads
\begin{equation}
\beta(\alpha_{\ell})= -2\alpha_{\ell}\left(1-\frac{t}{8\pi}\right)
+\frac{1}{16}\sum_{ij} '\alpha_i\alpha_j,
\label{b-alpha2}
\end{equation}
where the summation should take for those satisfying
$k_{\ell}=k_i\pm k_j$ ($i,j,\ell=1,\cdots,N$).

When $k_1$, $k_2$ and $k_3$ form an equilateral triangle ($M=3$), the
renormalization group equations for $\alpha_1$, $\alpha_2$ and $\alpha_3$
are closed within three equations.
When $k_1$, $k_2$, $\cdots$, $k_6$ form a regular tetrahedron ($M=6$),
we again have a closed set of equations for $\alpha_j$ ($j=1,2,\cdots,6$).
In the Wilsonian method, the same beta equation for $\alpha_{\ell}$ is
obtained which will be discussed in the next section.
In the Wilson renormalization method, however, a new vertex $k_i\mp k_j$ is
generated from any two vectors $k_i$ and $k_j$ except the case $k_i\cdot k_j=0$.

\vspace{0.5cm}
\begin{figure}[htbp]
\begin{center}
\includegraphics[height=7.0cm,angle=90]{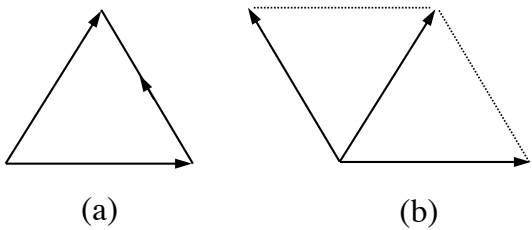}
\caption{
Triangles formed by wave vectors $k_i$, $k_j$ and $k_{\ell}$
for (a) $k_i\cdot k_j=1/2$ and (b) $k_i\cdot k_j=-1/2$.
}
\label{phase1}
\end{center}
\end{figure}

\subsection{Relation to the tachyon scattering amplitude in a bosonic string theory}

The two-vertex correction $J_{ij}$ is related to the tachyon scattering 
amplitude in a bosonic string theory.
The $n$-point scattering amplitude for tachyon scattering is given 
as\cite{kak88,kob69}
\begin{eqnarray}
A_n &:=& \int d\mu \int DX \exp\bigg[ -\frac{1}{4\pi\alpha'}\int
(\partial_zX_{\mu}\partial_{\bar{z}}X^{\mu})d^2z \nonumber\\
&& +i\sum_{i=1}^n k_{i\mu}X^{\mu} \bigg] \nonumber\\
&=& \int d\mu \prod_{1\leq i<j\leq n}|z_i-z_j|^{2\alpha' k_i\cdot k_j},
\end{eqnarray}
where the integration with the measure $d\mu$ is an integral over the
various $z_i$.
If we assume the correspondence
\begin{equation}
2\pi\alpha' = t,
\end{equation}
the $z_i$ dependence of the amplitude $A_2$ agrees with $J_{ij}$
where $J_{ij}\sim \int d^dr A_{ij}(r)^{-1}$ with
$A_{ij}(r)=r^{\alpha'k_i\cdot k_j}=r^{tk_i\cdot k_j/2\pi}$.
The vertex-vertex renormalization is given by the amplitude for
tachyon scattering.

\vspace{0.5cm}
\subsection{Renormalization group flow}

For an equilateral triangle configuration of $\{k_i\}$ with $M=3$ and $N=2$,
the equations read
\begin{eqnarray}
\mu\frac{\partial\alpha_1}{\partial\mu} &=& -2\alpha_1
\left(1-\frac{t}{8\pi}\right)+\frac{1}{16}\alpha_2\alpha_3, \nonumber\\
\mu\frac{\partial\alpha_2}{\partial\mu} &=& -2\alpha_2
\left(1-\frac{t}{8\pi}\right)+\frac{1}{16}\alpha_3\alpha_1, \nonumber\\
\mu\frac{\partial\alpha_3}{\partial\mu} &=& -2\alpha_3
\left(1-\frac{t}{8\pi}\right)+\frac{1}{16}\alpha_1\alpha_2,
\end{eqnarray}
and
\begin{equation}
\mu\frac{\partial t}{\partial\mu}= (d-2)t+\frac{C}{32M}t\sum_{i=1}^3\alpha_i^2
\end{equation}
We consider the simplified case where $\alpha_i=\alpha$ ($i=1,2,3$).
In this case the equations read
\begin{eqnarray}
\mu\frac{\partial\alpha}{\partial\mu} &=& -2\alpha
\left(1-\frac{t}{8\pi}\right)+\frac{1}{16}\alpha^2, \nonumber\\
\mu\frac{\partial t}{\partial\mu}&=& (d-2)t+\frac{C}{32}t\alpha^2.
\end{eqnarray}
In two dimensions $d=2$, the equations become
\begin{eqnarray}
\mu\frac{\partial\alpha}{\partial\mu} &=& 2\alpha v
+\frac{1}{16}\alpha^2, \nonumber\\
\mu\frac{\partial v}{\partial\mu}&=& \frac{C}{32}\alpha^2,
\end{eqnarray}
for $t=8\pi(1+v)$.
The renormalization group flow is shown in Fig. 2 for $\alpha >0$. The dotted line
indicates $\alpha=-32v$ where $\mu\partial\alpha/\partial\mu$ vanishes.
The asymptotic line as $\mu\rightarrow\infty$ is given by
$\alpha\sim b_{+}v$ with
\begin{equation}
b_{+}= \frac{1}{C}\left( 1+\sqrt{1+64C} \right),
\end{equation}
and $\alpha\sim b_{-}v$ with
\begin{equation}
b_{-}= \frac{1}{C}\left( 1-\sqrt{1+64C} \right).
\end{equation}
It is apparent from Fig. 2 that there is an asymmetry between
positive $v$ and negative $v$.  This is due to the two-vertex contribution.
There is also an asymmetry between $\alpha >0 $ and $\alpha <0$.
The flow for $\alpha <0$ is obtained just by extending straight lines
into the negative $\alpha$ region.

\vspace{0.5cm}
\begin{figure}[htbp]
\begin{center}
\includegraphics[height=6.0cm]{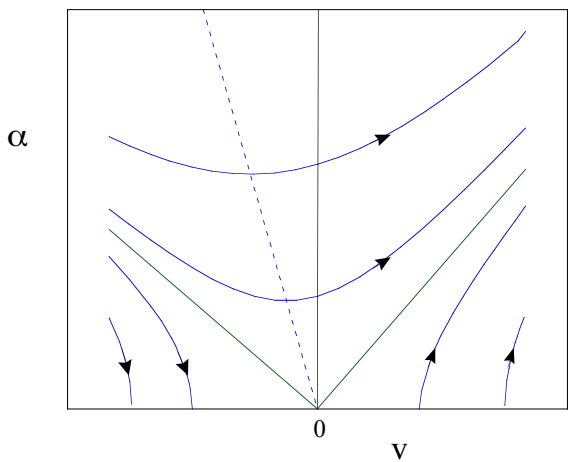}
\caption{
Renormalization group flow as $\mu\rightarrow\infty$ in the plane
of $\alpha$ and $v$.
}
\label{flow1}
\end{center}
\end{figure}

\section{Wilsonian renormalization group method}

We investigate the renormalization of the multi-vertex sine-Gordon
model by using the Wilsonian renormalization group method.
We obtain the same set of equations as that in the dimensional
regularization method.
The only difference is that two vertices satisfying $k_i\cdot k_j\neq 0$
generate a new vertex, while $k_i$ and $k_j$ should satisfy
$k_i\cdot k_j=\pm 1/2$ in the dimensional regularization method.

\subsection{Wilsonian renormalization procedure}

We write the action in the following form.
\begin{equation}
S = \int d^2x \bigg[ \frac{1}{2}(\partial_{\mu}\phi)^2
+ \sum_jg_j\cos(\beta k_j\cdot \phi)\bigg],
\end{equation}
where $g_j=\alpha_j/t$ and $\beta=\sqrt{t}$.
The field $\phi$ was scaled to $\beta\phi$.
We reduce the cutoff $\Lambda$ in the following way:
\begin{equation}
\Lambda\rightarrow \Lambda-d\Lambda= \Lambda-\Lambda d\ell=\Lambda e^{-d\ell}.
\end{equation}
The scalar field $\phi=(\phi_1,\cdots,\phi_N)$ is divided into
two parts as $\phi(x)=\phi_{1}(x)+\phi_{2}(x)$ with 
$\phi_{\ell}(x)= (\phi_{\ell 1},\cdots,\phi_{\ell N})$ ($\ell=1,2$) 
where
\begin{eqnarray}
\phi_{1j}(x)&=& \int_{0\leq |{\bf p}|\leq \Lambda-d\Lambda}\frac{d^2p}{(2\pi)^2}
e^{i{\bf p}\cdot x}\phi_j(x),\nonumber\\
\phi_{2j}(x)&=& \int_{\Lambda-d\Lambda\leq |{\bf p}|\leq \Lambda}\frac{d^2p}{(2\pi)^2}
e^{i{\bf p}\cdot x}\phi_j(x).
\end{eqnarray}
The action is written as
\begin{eqnarray}
S &=& \int d^2x\bigg[ \sum_{\ell=1}^2\frac{1}{2}(\partial_{\mu}\phi_{\ell})^2
 +\sum_jg_j\cos(\beta k_j\cdot (\phi_1+\phi_2))\bigg] \nonumber\\
&=& S_0(\phi_1)+S_0(\phi_2)+S_1(\phi_1,\phi_2),
\end{eqnarray}
where $S_1$ indicates the potential term.
Then the partition function is given by
\begin{eqnarray}
Z &=& \int {\cal D}\phi e^{-S}\nonumber\\
&=& \int {\cal D}\phi_1\exp\left(-S_0(\phi_1)+\sum_n^{\infty}\Gamma_n(\phi_1)\right),
\end{eqnarray}
where
\begin{equation}
\sum_{n=1}^{\infty}\Gamma_n(\phi_1)= {\Large \langle\langle} \sum_{n=0}^{\infty}
\frac{1}{n!}(-1)^nS_1^n {\Large \rangle\rangle}_{{\rm conn}},
\end{equation}
with
\begin{equation}
\langle\langle Q\rangle\rangle = \frac{1}{Z_2}\int {\cal D}\phi_2
e^{-S_0(\phi_2)}Q.
\end{equation}
$\langle\langle\cdot\rangle\rangle_{{\rm conn}}$ means keeping only connected
diagrams in $\langle\langle\cdot\rangle\rangle$.
$\Gamma_n$ ($n=1,2,\cdots$) represent contributions to the effective
action.

\subsection{Lowest order renormalization of $g_j$}

The lowest order contribution $\Gamma_1=-\langle\langle S_1\rangle\rangle$
reads
\begin{eqnarray}
\Gamma_1 &\simeq& -\sum_jg_j\int d^2x\cos(\beta k_j\cdot \phi_1)
\exp\left( -\frac{1}{2}\beta^2\langle\langle (k_j\cdot \phi_2)^2
\rangle\rangle \right)\nonumber\\
&=& -\sum_jg_j\exp\left(-\frac{1}{2}\beta^2 G_{jd\Lambda}(0)\right)
\int d^2x\cos(\beta k_j\cdot\phi_1),
\end{eqnarray} 
where the Green function $G_{jd\Lambda}$ is defined as
\begin{eqnarray}
G_{jd\Lambda}({\bf x}_1-{\bf x}_2) &=& \langle\langle \phi_{2j}
({\bf x}_1)\phi_{2j}({\bf x}_2)\rangle\rangle
\nonumber\\
&=& \frac{d\Lambda}{\Lambda}\frac{1}{2\pi}J_0(\Lambda|{\bf x}_1-{\bf x}_2|),
\end{eqnarray}
where $J_0$ is the zero-th Bessel function.
Up to this order, the action is renormalized to
\begin{eqnarray}
S_{\Lambda-d\Lambda}&=& S_0(\phi_1)-\Gamma_1 \nonumber\\
&=& \int d^2x \bigg[ \frac{1}{2}(\partial_{\mu}\phi_1)^2
+\sum_jg_j\left( 1-\frac{\beta^2}{2}G_{jd\Lambda}(0)\right) \nonumber\\
&&\times \cos(\beta k_j\cdot\phi)\bigg].
\end{eqnarray}
We perform the following scale transformation:
\begin{eqnarray}
{\bf x} &\rightarrow & {\bf x}'=e^{-d\ell}{\bf x}, \nonumber\\
{\bf p} &\rightarrow & {\bf p}'=e^{d\ell}{\bf p}, \nonumber\\
\phi_1({\bf p}) &\rightarrow & \tilde{\phi_1}({\bf p}')=\phi_1({\bf p})\zeta^{-1},
\label{scale}
\end{eqnarray}
where $\zeta$ is the scaling parameter for the field $\phi_1$.
In the real space we have
\begin{equation}
\phi_1({\bf x})= \zeta e^{-2d\ell}\tilde{\phi_1}({\bf x}').
\label{scale2}
\end{equation}
Then the effective action reads
\begin{eqnarray}
S_{\Lambda-d\Lambda}&=& \int d^2x'\bigg[
\zeta^2e^{-4d\ell}\frac{1}{2}(\partial_{\mu}'\tilde{\phi_1}({\bf x}'))^2
\nonumber\\
&+& \sum_jg_j e^{2d\ell}\left(1-\frac{\beta^2}{4\pi}\frac{d\Lambda}{\Lambda}\right)
\nonumber\\
&& \times\cos\left(\beta\zeta e^{-2d\ell}k_j\cdot\tilde{\phi_1}({\bf x}') 
\right)\bigg].
\end{eqnarray}
Here we put
\begin{equation}
\zeta^2 e^{-4d\ell}=1,
\end{equation}
so that we obtain
\begin{eqnarray}
S_{\Lambda}&=& \int d^2x'\bigg[
\frac{1}{2}(\partial_{\mu}'\tilde{\phi_1}({\bf x}'))^2
+ \sum_jg_j \left(1+2\frac{d\Lambda}{\Lambda}-\frac{\beta^2}{4\pi}
\frac{d\Lambda}{\Lambda}\right)
\nonumber\\
&& \times\cos\left(\beta k_j\cdot\tilde{\phi_1}({\bf x}') 
\right)\bigg].
\end{eqnarray}
This leads to the renormalized $g_{Rj}$ and $\beta_R$ as
\begin{eqnarray}
g_{Rj}&=& g_j+\left(2-\frac{\beta^2}{4\pi}\right)g_j\frac{d\Lambda}{\Lambda},
\\
\beta_R &=& \beta.
\end{eqnarray}
Then we have
\begin{eqnarray}
\Lambda\frac{dg_j}{d\Lambda} &=& \left(2-\frac{\beta^2}{4\pi}\right)g_j,
\nonumber\\
\Lambda\frac{d\beta}{d\Lambda} &=& 0.
\end{eqnarray}
Since $\beta^2=t$, these results agree with those obtained by the
dimensional regularization method in two dimensions.

\subsection{Multi-vertex contributions}

The second-order contribution to the effective action is
\begin{eqnarray}
\Gamma_2 &=& \frac{1}{2}\langle\langle S_1^2\rangle\rangle_{{\rm conn}} \nonumber\\
&=& \frac{1}{2}\sum_{ij}g_ig_j\int d^2x d^2x' \bigg[ \nonumber\\
&& \langle\langle \cos(\beta k_i\cdot\phi({\bf x}))
\cos(\beta k_j\cdot\phi({\bf x}'))\rangle\rangle \nonumber\\
&& -\langle\langle \cos(\beta k_i\cdot\phi({\bf x}))\rangle\rangle
\langle\langle \cos(\beta k_j\cdot\phi({\bf x}'))\rangle\rangle\bigg] .
\end{eqnarray}
We integrate out contributions with respect to $\phi_2$ variable.
For example, we use
\begin{eqnarray}
&& \langle\langle e^{i\beta(s k_i\cdot\phi_2({\bf x})
+s' k_j\cdot\phi_2({\bf x}')}\rangle\rangle
\nonumber\\
&=& \frac{1}{2}\beta^2\bigg[ (k_i^2+k_j^2)G_{d\Lambda}(0)
+2ss'k_i\cdot k_j G_{d\Lambda}({\bf x}-{\bf x}') \bigg],
\nonumber\\
\end{eqnarray}
where $s$ and $s'$ take $\pm 1$.
The second-order effective action $\Gamma_2$ is given as
\begin{eqnarray}
\Gamma_2 &=& \frac{1}{4}\sum_{ij}g_ig_j\exp\left( -\beta^2 G_{d\Lambda}(0)\right)
\int d^2x d^2x' \bigg[ \nonumber\\
&& \left(e^{-\beta^2 k_i\cdot k_j G_{d\Lambda}({\bf x}-{\bf x}')}-1\right)
\cos\left( \beta(k_i\cdot\phi_1({\bf x})+k_j\cdot\phi_1({\bf x}')) \right)
\nonumber\\
&+& \left(e^{\beta^2 k_i\cdot k_j G_{d\Lambda}({\bf x}-{\bf x}')}-1\right)
\cos ( \beta(k_i\cdot\phi_1({\bf x})
 -k_j\cdot\phi_1({\bf x}')) )\bigg].
\nonumber\\
\end{eqnarray}
When $k_i\cdot k_j>0$, the second term grows large for 
$|{\bf x}-{\bf x}'|\rightarrow 0$, while the first term becomes small.
When $k_i\cdot k_j<0$, the first term instead becomes large.  Hence we have
\begin{eqnarray}
\Gamma_2 &=& \frac{1}{4}\sum_{ij}g_ig_j\exp\left( -\beta^2 G_{d\Lambda}(0)\right)
\int d^2x d^2r \bigg[ \nonumber\\
&& \left(e^{\beta^2 |k_i\cdot k_j| G_{d\Lambda}({\bf r})}-1\right)
\cos\left( \beta(k_i\cdot\phi_1({\bf x})\mp k_j\cdot\phi_1({\bf x}+{\bf r})) 
\right)\bigg],
\nonumber\\
\end{eqnarray}
where $\mp$ takes $-$ when $k_i\cdot k_j>0$ and $+$ for $k_i\cdot k_j<0$.
Since the integrand is large when ${\bf r}$ is small, $\Gamma_2$ is written as
\begin{eqnarray}
\Gamma_2 &=& \frac{1}{4}\sum_{ij}g_ig_j\beta^2|k_i\cdot k_j|\int d^2r
G_{d\Lambda}(r) \nonumber\\
&\cdot& \int d^2x\cos(\beta(k_i\mp k_j)\phi_1({\bf x}))
\left(1-\frac{\beta^2}{2}({\bf r}\cdot\nabla(k_j\cdot\phi_1)^2)\right)
\nonumber\\
&\simeq& \frac{1}{4}\sum_{ij}g_ig_j\beta^2|k_i\cdot k_j|\int d^2r
G_{d\Lambda}(r)\nonumber\\
&\cdot& \int d^2x\cos(\beta(k_i\mp k_j)\phi_1({\bf x}))\nonumber\\
&-&\frac{1}{8}\sum_jg_j^2\beta^4\int d^2r r^2\frac{1}{2}G_{d\Lambda}(r)
\int d^2x(\partial_{\mu}(k_j\cdot\phi_1))^2,
\end{eqnarray}
where in the second term with derivative of $\phi_1$ we keep only
$k_i=k_j$ term since this term otherwise becomes small due to the 
oscillation of cosine function.
As discussed before, we use the approximation
$\sum_jg_j^2(\partial_{\mu}(k_j\cdot\phi_1))^2\simeq 
\langle g_j^2\rangle C(\partial_{\mu}\phi_1)^2$.

Then the effective action reads
\begin{eqnarray}
S_{\Lambda-d\Lambda}&=& S_0(\phi_1)-\Gamma_1-\Gamma_2 \nonumber\\
&=& \int d^2x\bigg[ \frac{1}{2}(\partial_{\mu}\phi_1)^2 
\left(1+\frac{A}{8}\beta^4\langle g_j^2\rangle\frac{d\Lambda}{\Lambda^5}\right)
\nonumber\\
&+& \sum_jg_j\left(1-\frac{\beta^2}{2}G_{d\Lambda}(0)\right)
\cos(\beta k_j\cdot\phi_1)\nonumber\\
&-& \frac{1}{4}B\sum_{ij}g_ig_j\beta^2|k_i\cdot k_j|\frac{d\Lambda}{\Lambda^3}
\cos(\beta (k_i\mp k_j)\phi_1({\bf x})) \bigg],
\nonumber\\
\end{eqnarray}
where $A$ ans $B$ are constants defined by
\begin{equation}
A=C\int_0^1drr^3J_0(r),~~~B= \int_0^1drrJ_0(r).
\end{equation}
We perform the scale transformation in eqs.(\ref{scale}) and (\ref{scale2})
where the parameter $\zeta$ is chosen as
\begin{equation}
\zeta^2 e^{-4d\ell}\left( 1+\frac{A}{8}\beta^4\langle g_j^2\rangle
\frac{d\Lambda}{\Lambda^5}
\right) =1.
\end{equation}
Then the renormalized action is given by
\begin{eqnarray}
S_{\Lambda}&=& \int d^2x\bigg[ \frac{1}{2}(\partial_{\mu}\tilde{\phi_1})^2
\nonumber\\
&+& \sum_jg_j\left( 1+2\frac{d\Lambda}{\Lambda}-\frac{\beta^2}{4\pi}
\frac{d\Lambda}{\Lambda} \right)\cos(\beta\zeta e^{-2d\ell}
k_j\cdot\tilde{\phi_1}({\bf x}))\nonumber\\
&-& \frac{B}{4}\beta^2\sum_{ij}g_ig_j|k_i\cdot k_j|\frac{d\Lambda}{\Lambda^3}
\cos(\beta\zeta e^{-2d\ell} (k_i\mp k_j)\cdot\tilde{\phi_1}({\bf x})) \bigg].
\nonumber\\
\end{eqnarray}
This results in the renormalization group equations as follows.
\begin{eqnarray}
\Lambda\frac{d\beta}{d\Lambda} &=& -\frac{A}{16\Lambda^4}\beta^5
\langle g_j^2\rangle,
\nonumber\\
\Lambda\frac{dg_j}{d\Lambda} &=& \left(2-\frac{\beta^2}{4\pi}\right)g_j
-\frac{B}{4\Lambda^2}\beta^2\sum_{i\ell}' g_ig_{\ell}|k_i\cdot k_{\ell}|,
\end{eqnarray}
where the summation is taken for $k_i$ and $k_{\ell}$ satisfying
$k_j=k_i\mp k_{\ell}$.

The resulting equations are consistent with those obtained using the
dimensional regularization.
Note that the sign is different because the derivative is calculated
in the descending direction $\Lambda\rightarrow \Lambda-d\Lambda$
in the Wilsonian method.
In the dimensional regularization method, the summation for $g_i$ and $g_{\ell}$
is restricted to $k_i$ and $k_{\ell}$ that satisfy $k_i\cdot k_{\ell}=\pm 1/2$.
In the Wilsonian method, this condition is relaxed and new vertex is
generated unles $k_i$ and $k_{\ell}$ are orthogonal.

\section{Generalized multi-vertex model}

\subsection{Renormalization of $\alpha_j$}

As shown in the evaluation of $\beta(t)$, the corrections to $t$
are dependent on momentum parameter $\{k_i\}$.
We examine this in this section.
We consider the generalized Lagrangian given as
\begin{equation}
\mathcal{L}= \sum_{\ell}\frac{Z_{\phi}}{2t_{\ell}\mu^{2-d}Z_{t_{\ell}}}
(\partial_{\mu}\phi_{\ell})^2
+\sum_j\frac{\mu^d\alpha_j Z_{\alpha_j}}{t_jZ_{t_j}}
\cos\left(\sqrt{Z_{\phi}}k_j\cdot\phi\right).
\end{equation} 
The potential term is renormalized to
\begin{equation}
\alpha_jZ_{\alpha_j}\exp\left( -\frac{1}{2}Z_{\phi}
\sum_{\ell}k_{j\ell}^2\langle\phi_{\ell}^2\rangle\right)
\cos\left(\sqrt{Z_{\phi}}k_j\cdot\phi\right),
\end{equation}
where
\begin{equation}
\langle\phi_{\ell}^2\rangle= \frac{t_{\ell}\mu^{2-d}Z_{t_{\ell}}}{Z_{\phi}}
\int\frac{d^dp}{(2\pi)^d}\frac{1}{p^2+m_0^2}
= -\frac{t_{\ell}\mu^{2-d}Z_{t_{\ell}}}{Z_{\phi}}\frac{1}{\epsilon}
\frac{\Omega_d}{(2\pi)^d}.
\end{equation}
Then the correction is written as
\begin{equation}
\alpha_jZ_{\alpha_j}\left( 1+\frac{1}{2\epsilon}\sum_{\ell}k_{j\ell}^2t_{\ell}
\mu^{2-d}Z_{t_{\ell}}\frac{\Omega_s}{(2\pi)^d}\right)
\cos\left(\sqrt{Z_{\phi}}k_j\cdot\phi\right).
\end{equation}
This results in
\begin{equation}
Z_{\alpha_j}= 1-\frac{1}{4\pi\epsilon}\sum_{\ell}k_{j\ell}^2t_{\ell}.
\end{equation}
Then we obtain
\begin{equation}
\mu\frac{\partial\alpha_j}{\partial\mu} = -2\alpha_j
\left( 1-\frac{1}{8\pi}\sum_{\ell}k_{j\ell}^2t_{\ell} \right).
\end{equation}
The fixed point of $\{t_{\ell}\}$ is obtained as a zero of this equation.
For an equilateral triangle where $N=2$ and $M=3$, we can choose $\{k_j\}$ as
\begin{equation}
k_1=(1,0),~~~ k_2=(1/2,\sqrt{3}/2),~~~ k_3=(-1/2,\sqrt{3}/2).
\end{equation}
The critical value of $t_{\ell}$ is obtained as
\begin{equation}
t_{1c}=t_{2c}=t_{3c}=8\pi.
\end{equation}
For $N=3$, we can consider a regular tetrahedron with $M=6$, $\{k_j\}$
are set as
\begin{eqnarray}
k_1&=&(1,0,0),~ k_2=(1/2,\sqrt{3}/2,0),~ k_3=(-1/2,\sqrt{3}/2,0),\nonumber\\ 
k_4&=&(1/2,1/2\sqrt{3},\sqrt{2/3}),~~ k_5=(1/2,-1/2\sqrt{3},\sqrt{2/3})
\nonumber\\
k_6&=&(0,-1/\sqrt{3},\sqrt{2/3}).
\end{eqnarray}
In this case, the fixed point of $\{t_{\ell}\}$ is also given by
$t_{1c}=t_{2c}=\cdots = t_{6c}=8\pi$.

\subsection{Renormalization of $t_{\ell}$}

From the second-order perturbation, there appears the term that
renormalizes the kinetic term as shown in section III.
We use the following approximation here:
\begin{eqnarray}
&&\cos\left(\sqrt{Z_{\phi}}k_j\cdot (\phi({\bf x})-\phi({\bf x}+{\bf r}))\right)
\nonumber\\
&& ~~~= \cos\left( \sqrt{Z_{\phi}}r_{\mu}\partial_{\mu}(k_j\cdot\phi({\bf x}))
-\cdots\right) \nonumber\\
&& ~~~ = 1-\frac{1}{2}Z_{\phi}\left( r_{\mu}\partial_{\mu}(k_j\cdot\phi({\bf x}))
\right)^2+\cdots \nonumber\\
&& ~~~ = 1-\frac{1}{2}Z_{\phi}r_{\mu}r_{\nu}\sum_{\ell m}k_{j\ell}k_{jm}
\partial_{\mu}\phi_{\ell}\partial_{\nu}\phi_m +\cdots.
\end{eqnarray}
We keep the diagonal terms $(\partial_{\mu}\phi_{\ell})^2$, and then $I_{1a}^{(2)}$
in section III becomes
\begin{equation}
I_{1a}^{(2)}\simeq -\frac{1}{32\epsilon}\sum_{j\ell}
\frac{\mu^{d-2}Z_{\phi}}{2t_jZ_{\alpha_j}}\mu^{d+2}a^4\alpha_j^2
k_{j\ell}^2(\partial_{\mu}\phi_{\ell})^2,
\end{equation}
where we put $t_{\ell}=8\pi (1+v_{\ell})$ and neglect the term of the order 
of $v_{\ell}$.  Then the kinetic term is renormalized into
\begin{equation}
\sum_{\ell}\frac{Z_{\phi}}{2t_{\ell}\mu^{2-d}Z_{t_{\ell}}}\bigg[
1-\frac{1}{32\epsilon}\mu^{d+2}a^4\sum_j\alpha_j^2k_{j\ell}^2\bigg]
(\partial_{\mu}\phi_{\ell})^2.
\end{equation}
This leads to
\begin{equation}
Z_{t_{\ell}}= 1-\frac{1}{32\epsilon}\mu^{d+2}a^4\sum_j\alpha_j^2k_{j\ell}^2.
\end{equation}
Then we obtain
\begin{equation}
\mu\frac{\partial t_{\ell}}{\partial\mu}= (d-2)t_{\ell}
+\frac{1}{32}t_{\ell}\sum_j\alpha_j^2k_{j\ell}^2.
\end{equation}
For $N=2$ and $M=3$, we use $\{k_j\}$ for an equilateral triangle, the
equations for $t_1$ and $t_2$ read
\begin{eqnarray}
\mu\frac{\partial t_1}{\partial\mu}= (d-2)t_1
+\frac{1}{32}t_1\left( \alpha_1^2+\frac{1}{4}\alpha_2^2+\frac{1}{4}\alpha_3^2
\right),\\
\mu\frac{\partial t_2}{\partial\mu}= (d-2)t_2
+\frac{1}{32}t_2\left( \frac{3}{4}\alpha_2^2+\frac{3}{4}\alpha_3^2
\right).
\end{eqnarray}
This is the result for the generalized multi-vertex sine-Gordon model.
The qualitative property is the same as that obtained for that in
section III.
When $\alpha\equiv \alpha_1\sim \alpha_2\sim \alpha_3$, we have
\begin{equation}
\mu\frac{\partial t_1}{\partial\mu}= (d-2)t_1+\frac{C}{32}t_1\alpha^2,
\end{equation}
with $C=3/2$.
This agrees with the previous result.

\subsection{Multi-vertex contribution to $\alpha_j$}
  
For the generalized model, $I_{1b}^{(2)}$ in section III becomes
\begin{eqnarray}
I_{1b}^{(2)} &=& -\frac{1}{4}\sum_{j\neq i}
\frac{\mu^d\alpha_jZ_{\alpha_j}}{t_jZ_{t_j}}
\frac{\mu^d\alpha_jZ_{\alpha_i}}{t_jZ_{t_i}}\int d^dx d^dr \nonumber\\
&& \exp\Big[ -\frac{Z_{\phi}}{2}\sum_{\ell}(k_{j\ell}^2+k_{i\ell}^2)
\langle\phi_{\ell}^2\rangle + \sum_{\ell}k_{j\ell}k_{i\ell}t_{\ell}
\nonumber\\
&&\times \mu^{2-d}Z_{t_{\ell}}\frac{1}{2\pi}K_0(m_0r)\Big] \nonumber\\
&& \cdot \cos\left( \sqrt{Z_{\phi}}(k_i\cdot\phi(x)-k_j\cdot\phi(x+{\bf r})) 
\right).
\end{eqnarray}
The integral with respect to $r$ becomes
\begin{eqnarray}
J_{ij}&:=& \int d^dr \exp\left( \sum_{\ell}k_{j\ell}k_{i\ell}t_{\ell}
\mu^{2-d}Z_{t_{\ell}}\frac{1}{2\pi}K_0(m_0r) \right) \nonumber\\
&\simeq& \Omega_d\int_0^{\infty}dr r^{d-1}
\left(\frac{1}{cm_0^2(r^2+a^2)}\right)^{\sum_{\ell}k_{j\ell}k_{i\ell}t_{\ell}/4\pi}.
\end{eqnarray}
We consider the region near the fixed point $t_{\ell}=8\pi(1+v_{\ell})$, 
where $J_{ij}$ is estimated as
\begin{equation}
J_{ij}= -\Omega_d(cm_0^2)^{-1}\frac{1}{\epsilon}+O(v),
\end{equation}
when $k_i\cdot k_j=1/2$.
This indicates that
\begin{equation}
I_{1b}^{(2)}\simeq \frac{1}{16\epsilon}\frac{cm_0^2}{8\pi}\sum_{i\neq j}
\alpha_j\alpha_i\mu^{2d}a^4\int d^dx
\cos\left(\sqrt{Z_{\phi}}(k_i-k_j)\cdot\phi(x)\right).
\end{equation}
The potential term with two-vertex correction is obtained as
\begin{eqnarray}
&&\sum_{\ell}\frac{\mu^d\alpha_{\ell}Z_{\alpha_{\ell}}}{t_{\ell}Z_{t_{\ell}}}
\bigg[ 1+\frac{1}{4\pi\epsilon}\sum_m k_{\ell m}^2t_m \nonumber\\
&&~~  +\frac{1}{16\epsilon}\sum_{i\neq j}'\frac{\alpha_j\alpha_i}{\alpha_{\ell}}
\mu^d cm_0^2a^4 \bigg]
\cos\left(\sqrt{Z_{\phi}}k_{\ell}\cdot\phi\right),
\end{eqnarray}
where $\sum_{i\neq j}'$ indicates the summation under the condition 
that $k_i\pm k_j=k_{\ell}$.
Then we choose $Z_{\alpha_{\ell}}$ as
\begin{equation}
Z_{\alpha_{\ell}}= 1-\frac{1}{4\pi\epsilon}\sum_{m}k_{\ell m}^2t_{m}
-\frac{1}{16\epsilon}\sum_{i\neq j}'\frac{\alpha_i\alpha_j}{\alpha_{\ell}}
\mu^d cm_0^2a^4.
\end{equation}
The beta function up to the second order of $\alpha$ is given as
\begin{equation}
\mu\frac{\partial\alpha_{\ell}}{\partial\mu}=
-2\alpha_{\ell}\left(1-\frac{1}{8\pi}\sum_mk_{\ell m}^2t_m \right)
+\frac{1}{16}\sum_{i\neq j}'\alpha_i\alpha_j,
\end{equation}
where we set $cm_0^2a^2=1$.

\section{Summary}

We investigated the multi-vertex sine-Gordon model on the basis of
the renormalization group theory.
We employ the dimensional regularization method and the Wilsonian
renormalization group method.  Two results are consistent each other.
The generalized sine-Gordon model contains multiple cosine (vertex) potentials
labelled by momentum parameters $\{k_j\}_{j=1,\cdots,M}$.
The vertex-vertex scattering amplitude is given by tachyon scattering amplitude.
A new vertex $k_{\ell}$ is generated from two vertex interactions $k_i$ and $k_j$,
 and they are
closed when momentum parameters $\{k_j\}$ satisfy the triangle condition
that $k_i\pm k_j= k_{\ell}$.
When $k_i$ and $k_j$ are orthogonal, a new vertex is not generated.
The condition $k_i\cdot k_j=\pm 1/2$ is required in the dimensional
regularization method.

For two-component scalar field ($N=2$), $\{k_j\}$ should form a triangle
(Wilson method) or
an equilateral triangle (dimensional regularization) for $M=3$.
For three-component scalar field ($N=3$), a regular tetrahedron form
a closed system for $M=6$.
For these structures, the fixed point of $\{t_j\}$ is given by
$t_1=t_2=\cdots =t_M$.
A regular octahedron is also possible where there are six independent $k_j$s and
thus $M=6$.
For an equilateral triangle, regular tetrahedron and regular octahedron,
we have $\sum_{j}k_{j\ell}^2=C(M)$ for $\ell=1,\cdots,N$ where we impose
the normalization $\sum_{\ell}k_{j\ell}^2=1$.
We expect that there exist crystal structures in higher dimensions
$N\geq 3$ satisfying $\sum_{j}k_{j\ell}^2={\rm const.}$ for any $\ell$.

The beta function of $\alpha_{\ell}$ is generalized to include the
product $\alpha_i\alpha_j$ for which $k_i\pm k_j=k_{\ell}$ is satisfied.
This term is a non-trivial contribution compared to the conventional
sine-Gordon model.
The beta function of $t_{\ell}$ has also contributions proportional to
$\alpha_j^2$.  These terms are positive and thus do not change
the renormalization group flow of $t_{\ell}$.
The additional terms to $\beta(\alpha_{\ell})$ change the flow of
$(\alpha_{\ell},t_j)$ qualitatively.

This work was supported by a Grant-in-Aid for Scientific
Research from the Ministry of Education, Culture, Sports, Science and
Technology of Japan (Grant No. 17K05559).

\end{document}